\documentclass[twocolumn]{aastex631}

\usepackage{array}
\usepackage{makecell}

\newcommand{\ebvgas}{$E(B-V)_{\text{gas}}$}
\newcommand{\one}{~\textsc{i}}
\newcommand{\ii}{~\textsc{ii}}
\newcommand{\iii}{~\textsc{iii}}

\newcommand{\ott}{O$_{32}$}
\newcommand{\rtt}{R$_{23}$}
\newcommand{\mstar}{M$_*$}
\newcommand{\othb}{[O~\textsc{iii}]/H$\beta$}
\newcommand{\ntha}{[N~\textsc{ii}]/H$\alpha$}
\newcommand{\stha}{[S~\textsc{ii}]/H$\alpha$}
\newcommand{\oiha}{[O~\textsc{i}]/H$\alpha$}

\begin{document}

\title{Excitation and Ionization Properties of Star-forming Galaxies at $z=2.0-9.3$ with {\it JWST}/NIRSpec}

\author[0000-0003-4792-9119]{Ryan L. Sanders}\altaffiliation{NHFP Hubble Fellow}\affiliation{Department of Physics and Astronomy, University of California, Davis, One Shields Ave, Davis, CA 95616, USA}

\email{email: rlsand@ucdavis.edu}

\author[0000-0003-3509-4855]{Alice E. Shapley}\affiliation{Department of Physics \& Astronomy, University of California, Los Angeles, 430 Portola Plaza, Los Angeles, CA 90095, USA}

\author{Michael W. Topping}\affiliation{Steward Observatory, University of Arizona, 933 N Cherry Avenue, Tucson, AZ 85721, USA}

\author[0000-0001-9687-4973]{Naveen A. Reddy}\affiliation{Department of Physics \& Astronomy, University of California, Riverside, 900 University Avenue, Riverside, CA 92521, USA}

\author[0000-0003-2680-005X]{Gabriel B. Brammer}\affiliation{Cosmic Dawn Center (DAWN), Denmark}\affiliation{Niels Bohr Institute, University of Copenhagen, Lyngbyvej 2, DK2100 Copenhagen \O, Denmark}

\begin{abstract}
We utilize medium-resolution {\it JWST}/NIRSpec observations of 164 galaxies at $z=2.0-9.3$ from the Cosmic Evolution Early Release Science (CEERS) survey to investigate the evolution of the excitation and ionization properties of galaxies at high redshifts.
Our results represent the first statistical constraints on the evolution of the [O\iii]/H$\beta$ vs.\ [N\ii]/H$\alpha$, [S\ii]/H$\alpha$, and [O\one]/H$\alpha$ ``BPT'' diagrams at $z>2.7$, and the first analysis of the \ott\ vs.\ \rtt\ diagram at $z>4$ with a large sample.
We divide the sample into five redshift bins containing $30-40$ galaxies each.
The subsamples at $z\sim2.3$, $z\sim3.3$, and $z\sim4.5$ are representative of the main-sequence star-forming galaxy population at these redshifts, while the $z\sim5.6$ and $z\sim7.5$ samples are likely biased toward high specific star-formation rate due to selection effects.
Using composite spectra, we find that each subsample at $z=2.0-6.5$ falls on the same excitation sequence in the [N\ii] and [S\ii] BPT diagrams and the \ott-\rtt\ diagram on average, offset from the sequences followed by $z=0$ H\ii\ regions in the same diagrams.
The direction of these offsets are consistent with high-redshift star-forming galaxies uniformly having harder ionizing spectra than typical local galaxies at fixed nebular metallicity.
The similarity of the average line ratios suggests that the ionization conditions of the interstellar medium do not strongly evolve between $z\sim2$ and $z\sim6$.
Overall, the rest-optical line ratios suggest the $z=2.7-9.3$ CEERS/NIRSpec galaxies at log($M_*/M_{\odot})\sim7.5-10$ have high degrees of ionization and moderately low oxygen abundances ($\sim0.1-0.3~Z_{\odot}$), but are not extremely metal poor ($<0.1~Z_{\odot}$) even at $z>6.5$.
\end{abstract}

\section{Introduction} \label{sec:intro}

Bright rest-frame optical emission lines are powerful tracers of the chemical and ionization state of the interstellar medium (ISM) of galaxies.
The relative strengths of nebular emission lines of different metal ions and hydrogen recombination lines are determined by the physical properties of the ionizing source and ionized gas, including the shape of the ionizing spectrum, the ionization parameter (a combination of the intensity of the radiation field and the geometry of the gas relative to the ionizing source), and gas-phase metallicity.

In the local universe, many diagnostic line ratio diagrams have been established to probe the physical conditions of the ionized ISM.
Diagrams plotting [O\iii]$\lambda$5008/H$\beta$ vs.\ [N\ii]$\lambda$6585/H$\alpha$, [S\ii]$\lambda\lambda$6718,6733/H$\alpha$, and [O\one]$\lambda$6302/H$\alpha$ have been used to differentiate normal H\ii\ regions and star-forming galaxies ionized by massive stars from sources with other ionization mechanisms, including active galactic nuclei (AGNs) \citep{bpt1981,vo1987,kau2003,kew2006}.
We refer to these line-ratio diagrams as the [N\ii], [S\ii], and [O\one] ``BPT'' diagrams, respectively.
Within the [N\ii] BPT diagram, $z\sim0$ star-forming galaxies follow a narrow sequence from high-\othb\ and low-\ntha\ to low-\othb\ and high \ntha.
This sequence represents a progression of increasing metallicity and decreasing levels of excitation and ionization, and also constitutes a sequence of increasing stellar mass (\mstar) reflecting the mass-metallicity relation \citep{tre2004,cur2020}.
Local star-forming galaxies and H\ii\ regions follow similar well-defined sequences in the [S\ii] and [O\one] BPT diagrams.

The location of the star-forming galaxy sequence in the [N\ii] BPT diagram evolves with redshift toward higher \ntha\ at fixed \othb, or equivalently higher \othb\ at fixed \ntha, between $z\sim0$ and $z\sim2.3$ \citep{ste2014,sha2015,sha2019,san2016,str2017,str2018,kas2017,kas2019fmos}.
This offset from the local sequence appears to steadily increase with redshift over this redshift range \citep{sha2019}.
The distinct line ratio excitation sequences followed by high- and low-redshift galaxies implies that at least some of the physical conditions within H\ii\ regions that regulate nebular line production evolve with redshift.
Recent work has converged on the evolution of the ionizing spectral shape at fixed nebular metallicity as the primary driver of the $z\sim2$ [N\ii] BPT diagram offset,
with any change in ionization parameter, electron density, and/or N/O abundance ratio playing only minor roles \citep{ste2016,str2018,sha2019,top2020a,top2020b,san2020,run2021}.
The different set of ionization properties at low and high redshifts has important implications for diagnostic calibrations used to translate line ratios to physical properties such as gas-phase metallicity that are of key importance for understanding galaxy evolution.
Because the ionization conditions differ at $z\sim0$ and $z\sim2$, deriving accurate metallicities for high-redshift samples requires a different set of metallicity calibrations from those based on typical $z\sim0$ H\ii\ regions and star-forming galaxies \citep[e.g.,][]{san2020}.

Another important diagnostic diagram relates the \ott\ ([O\iii]$\lambda$5008/[O\ii]$\lambda$3728) and \rtt\ (([O\iii]$\lambda\lambda$4960,5008+[O\ii]$\lambda$3728)/H$\beta$) line ratios.
\ott\ serves as a relatively direct proxy of the degree of ionization while \rtt\ is primarily sensitive to gas-phase metallicity.
The \ott-\rtt\ diagram has been studied out to $z\sim3.3$, with high-redshift galaxies generally displaying higher \ott\ and \rtt\ values on average relative to local galaxies \citep[e.g.,][]{nak2014,san2016,str2017,ono2016}, associated with their lower metallicities due to the evolving mass-metallicity relation \citep{san2021}.
In this diagram, a harder ionizing spectrum pushes the maximum \rtt\ ratio to higher values and generally increases \rtt\ at fixed \ott\ \citep{ste2016}.
Star-forming galaxies also follow a metallicity sequence in the \ott-\rtt\ diagram from low-\rtt\ and low-\ott\ at high metallicity to high-\rtt\ and high-\ott\ at low metallicity, though \rtt\ turns over and begins decreasing at very low metallicity \citep[$\lesssim0.1~Z_{\odot}$; e.g.,][]{kew2019}.
These line ratios are commonly used as tracers of ISM chemical abundance \citep[e.g.,][]{mai2008,tro2014,ono2016,san2021}.

Before the advent of {\it JWST}, ground-based observations of rest-optical emission lines at high redshift were limited to $z<2.7$ for the BPT diagrams, and $z<3.8$ for the \ott-\rtt\ diagram, where the necessary lines fall bluewards of the red edge of the near-infrared $K$-band at 2.4~$\mu$m.
Spectroscopic instruments onboard {\it JWST} now provide the ability to measure rest-optical lines at much higher redshifts, with the NIRSpec instrument covering H$\alpha$, [N\ii] and [S\ii] out to $z=6.5$, and [O\iii] and H$\beta$ out to $z=9.3$.
Recent studies of {\it JWST} spectra have provided the first glimpses into the metallicity and ionization properties of the ISM at $z=4-8$ based on rest-optical emission lines \citep[e.g.,][]{cur2023,bri2022,sch2022,are2022,tay2022,wan2022,tru2022,tac2022}, but these studies are based on small numbers of targets and their relation to the typical galaxy population is unclear.

In this paper, we use {\it JWST}/NIRSpec spectroscopy of a sample of 164 galaxies at $z=2.0-9.3$ to study the excitation and ionization properties of star-forming galaxies over cosmic time in diagnostic line ratio diagrams.
This is the first such study to employ a large statistical sample of galaxies, with particularly novel investigations of the BPT diagrams at $z>2.7$ and the \ott-\rtt\ diagram at $z>4$.
This paper is organized as follows. 
In Section~\ref{sec:obs}, we describe the observations, data reduction, measurements, and derived quantities.
We characterize the properties of the sample and the position of star-forming galaxies in diagnostic line-ratio diagrams in Section~\ref{sec:results}, investigating the BPT diagrams in Sec.~\ref{sec:bpt} and the \ott-\rtt\ diagram in Sec.~\ref{sec:r23o32}.
We discuss the implications of our findings in Section~\ref{sec:discussion}.

Throughout this paper, we adopt a \citet{cha2003} initial mass function (IMF), \citet{asp2021} solar abundance, and a cosmology described by $H_0=70\mbox{ km  s}^{-1}\mbox{ Mpc}^{-1}$, $\Omega_m=0.30$, and
$\Omega_{\Lambda}=0.7$.
We provide vacuum rest-frame emission-line wavelengths.
The line ratios analyzed in this work are defined as follows:
\begin{equation}
[\text{O}~\textsc{iii}]/\text{H}\beta = [\text{O}~\textsc{iii}]\lambda5008/\text{H}\beta
\end{equation}
\begin{equation}
[\text{N}~\textsc{ii}]/\text{H}\alpha = [\text{N}~\textsc{ii}]\lambda6585/\text{H}\alpha
\end{equation}
\begin{equation}
[\text{S}~\textsc{ii}]/\text{H}\alpha = [\text{S}~\textsc{ii}]\lambda\lambda6718,6733/\text{H}\alpha
\end{equation}
\begin{equation}
[\text{O}~\textsc{i}]/\text{H}\alpha = [\text{O}~\textsc{i}]\lambda6302/\text{H}\alpha
\end{equation}
\begin{equation}
\text{O}_{32} = [\text{O}~\textsc{iii}]\lambda5008/[\text{O}~\textsc{ii}]\lambda3728
\end{equation}
\begin{equation}
\text{R}_{23} = \frac{[\text{O}~\textsc{iii}]\lambda\lambda4960,5008+[\text{O}~\textsc{ii}]\lambda3728}{\text{H}\beta}
\end{equation}

\section{Observations and measurements}\label{sec:obs}

We use public data from {\it JWST}/NIRSpec Micro-Shutter Array (MSA) observations obtained as a part
 of the Cosmic Evolution Early Release Science (CEERS) survey \citep[Program ID: 1345][]{fin2022a,fin2022b}.
Medium-resolution spectroscopy was obtained using the G140M/F100LP, G235M/F170LP, and G395/F290LP
 grating and filter combinations for six pointings in the AEGIS field, providing a spectral
  resolution of $R\sim1000$ spanning wavelengths of $1-5~\mu$m.
Each grating configuration was observed for 3107 sec of on-source integration per pointing, divided into three
 exposures of 14 groups, with the NRSIRS2 readout mode.
Three MSA shutters were opened on each target to form slitlets of approximately $1.5''\times0.2''$,
 and a 3-point nod pattern was adopted.
Each multi-object slitmask contained between 52 and 55 primary targets.
In total, 318 unique sources were targeted, with three galaxies observed twice on two different masks.

\subsection{Data Reduction}

The data in all grating configurations were reduced using the following method.
The individual uncalibrated detector images were first passed through {\it JWST} \texttt{calwebb\_detector}
 pipeline\footnote{\url{https://jwst-pipeline.readthedocs.io/en/latest/index.html}}, in which
  bias and dark current were removed and saturated pixels and cosmic ray artifacts
 (e.g., ``showers'' and ``snowballs'') were masked.
The resulting images were then corrected for striping by subtracting an estimate of the 1/$f$ noise
 in each image.
The 2D spectrum for each slitlet was then cut out.
A flat-field correction, background subtraction based on dithered exposures,
photometric calibration, and wavelength solution were applied using the latest
 calibration reference data system (CRDS) context (\texttt{jwst\_1027.pmap}).
The cutout 2D spectrum for each target was rectified and interpolated onto a wavelength grid that was
 common for all observations in a particular grating configuration.
The calibrated 2D cutouts for each exposure were then combined based on the three-point dither pattern
 excluding pixels that were masked.
 Error spectra were derived
 by combining the variances from Poisson noise, read noise, and flat-fielding, and the variance
between each individual exposure.
This process yielded viable 2D science and error spectra for 310 targets in each grating and detector combination. 
For each grating configuration, the fully calibrated and rectified 2D spectra from both detectors
were combined into a single 2D spectrum spanning the full wavelength range including the chip gap.

One-dimensional science and error spectra were extracted from the rectified 2D spectra using an optimal extraction \citep{horne1986}.
The spatial profile was obtained by manually identifying wavelength ranges in the 2D spectrum containing high-S/N emission lines
 when present or detected continuum otherwise and summing the corresponding columns of the 2D spectrum.
A Gaussian profile was then fit to the resulting spatial profile, and the best-fit Gaussian model was used to calculate the weights
 in the optimal extraction scheme.
The extraction was limited to the range of cross-dispersion pixels around the peak where the spatial profile was positive.
The extraction windows were typically 5 pixels or 0.5$''$ in extent along the slit, approximately the size of one microshutter.
During the extraction, wavelength ranges in which the extracted 1D spectrum was affected by unremoved cosmic rays, artifacts,
 or emission from neighboring serendipitously detected sources were manually identified and masked.
The chip gap was also automatically identified and masked.
All masked wavelength ranges were considered to have no coverage during the subsequent emission-line fitting.
For many targets, spatial profiles could only be obtained for one or two grating configurations in which lines or continuum was detected,
 while the other gratings showed no significant detections from which to construct the profile.
In such cases, the mean centroid, FWHM, and extraction window of the manually extracted grating(s) were used to automatically
 extract spectra from the gratings that lacked detections.
1D spectra were not extracted for targets for which there were no visible detections in any grating configuration.
Out of 310 total targets, 1D spectra were extracted for 252 sources.

A wavelength-dependent slit-loss correction was then applied to the 1D spectra,
described in detail in Reddy et al.\ (2023, in prep.).
The intrinsic morphology of each target was estimated from {\it JWST}/NIRCam F115W imaging when available,
  or from a S\'{e}rsic profile fit to HST/WFC3 F160W imaging otherwise \citep{vdw2014}.
If a target lacked robust shape constraints from {\it JWST} and HST imaging, then it was assumed to be a point source.
At each wavelength, the fraction of total light falling inside an area defined by the width of the slit and
 the vertical extraction window was then estimated using the position of the target in the slitlet
 and the intrinsic morphology convolved with the wavelength-dependent {\it JWST}/NIRSpec point spread function.
The 1D spectrum was then divided by this fraction as a function of wavelength.

The final flux calibration was achieved by scaling the 1D science spectra to match the photometric SEDs in the following way.
The slit-loss corrected 1D spectra were passed through the filter transmission curves for the set of photometric bandpasses available for each target
 to produce synthetic photometric flux densities and errors.
The ratio of the image-based and synthetic flux densities was calculated for each filter in which both the synthetic and image-based
 photometric measurements had S/N$>$5.
If the number of filters meeting this requirement was greater than or equal to 3, then the 1D spectra and error spectra in all
 grating configurations were scaled by the median of the individual ratios to achieve the final flux calibration.
For the 109 targets that did not meet this criterion, no scale factor was applied.
For the remaining 143 targets, the median scale factor was 0.997 with a standard deviation of 0.23~dex,
 implying that the absolute flux calibration is robust and that flux calibration of targets for which a scale factor
  was not applied should not be biased on average.

\subsection{Emission Line Measurements}\label{sec:linefitting}

Emission line fluxes were measured from the 1D science spectra using the following method.
For each target, the redshift was first measured using the best-fit centroid from a single Gaussian fit to the
 line with the highest signal-to-noise ratio, usually [O~\textsc{iii}]$\lambda$5008 (57\%) or H$\alpha$ (36\%).
The intrinsic velocity FWHM of the highest signal-to-noise line (corrected for instrumental resolution) was also calculated from
 the best-fit Gaussian profile.
The instrumental resolution as a function of wavelength in each NIRSpec grating configuration was obtained from the dispersion curves
 distributed in the {\it JWST} User Documentation.\footnote{\url{https://jwst-docs.stsci.edu/jwst-near-infrared-spectrograph/nirspec-instrumentation/nirspec-dispersers-and-filters}}
For the extracted 1D science spectrum in each grating configuration, lines that fell within the covered wavelength
 range were then fit with Gaussian profiles for which the centroid was restricted to be within 50~km~s$^{-1}$ of
 the expected observed wavelength based on the measured redshift, while the FWHM was restricted to be within
 20\% of the expected FWHM in \AA\ calculated by convolving the intrinsic velocity FWHM measured above with the
 instrumental resolution at the expected line centroid and converting from km~s$^{-1}$ to \AA.
Adjacent lines separated by $\Delta\lambda$ at wavelength $\lambda$ with $\Delta\lambda/\lambda<0.01$ (e.g., [N~\textsc{ii}]$\lambda$6550, H$\alpha$, and [N~\textsc{ii}]$\lambda$6585)
were fit simultaneously with multiple Gaussians, while more widely separated lines were fit individually.
Closely spaced lines that are blended and unresolvable at $R\sim1,000$
 (e.g., [O~\textsc{ii}]$\lambda\lambda$3727,3730; [Ne~\textsc{iii}]$\lambda$3970 and H$\epsilon$; He~\textsc{i}~$\lambda$3890 and H$\zeta$)
 were fit with a single Gaussian for which the FWHM restriction is relaxed to 50\%.
The continuum model was taken to be the best-fit SED model (see Sec.~\ref{sec:sed} below), where the only free parameter is an additive offset.
Using the best-fit SED model as the continuum has the advantage of self-consistently accounting for stellar Balmer absorption such that
 the measured H recombination line fluxes are robust.

Neighboring NIRSpec medium-resolution gratings have overlapping wavelength coverage such that the same
 emission line was measured in two gratings for 90 unique targets.
Line fluxes measured in two gratings showed good agreement, with a median offset of 0.02~dex and an intrinsic scatter of 0.08~dex,
 suggesting that the relative flux calibration between grating configurations is robust on average.
The final measured line flux was taken to be the inverse-variance weighted mean of the fluxes measured in each grating when
 a line was covered in two configurations.

\subsection{Photometry and SED Fitting}\label{sec:sed}

Stellar population parameters including stellar mass are inferred by modeling the spectral energy distribution
 of each target using extensive multi-wavelength photometric catalogs.
For 99 targets that were covered by CEERS {\it JWST}/NIRCam imaging taken in June 2022 \citep{fin2022b},
 we used the catalog assembled by G. Brammer\footnote{\url{https://s3.amazonaws.com/grizli-v2/JwstMosaics/v4/index.html}}.
 This catalog includes aperture-matched measurements in 7 {\it JWST}/NIRCam filters
 (F115W, F150W, F200W, F277W, F356W, F410M, and F444W) and 7 HST/ACS and WFC3 filters
 (F435W, F606W, F814W, F105W, F125W, F140W, and F160W), spanning
 $\sim0.4-5~\mu$m in the observed frame.
A further 185 targets have measurements in the 3D-HST photometric catalogs
 \citep{mom2016,ske2014} spanning observed-frame $0.4-8~\mu$m that we used for SED fitting.
The remaining 35 CEERS medium-resolution NIRSpec targets had no coverage
 in the Brammer catalog and no reliable counterpart in the 3D-HST catalog, such that stellar
 population properties could not be derived for these targets.
Of the 231 galaxies with robust NIRSpec spectroscopic redshifts,, robust photometric SED information was available for a total of 210.

The SED fitting code FAST \citep{kri2009} was used to fit the flexible stellar population synthesis
 models \citep[FSPS;][]{con2009} to the measured photometry.
We assumed a \citet{cha2003} IMF, delayted-$\tau$ star-formation histories
 of the form $\mbox{SFR}\propto t\times e^{-t/\tau}$,
and two possible combinations of
 stellar metallicity and dust attenuation curve to reflect the chemical maturity of
 galaxies in different redshift and stellar mass regimes,
 following \citet{red2018} and \citet{du2018}.
As described in \citet{sha2023},
 we assumed $Z_*=0.02$ (1.4~$Z_{\odot}$) and the \citet{cal2000} attenuation curve for all galaxies
 at $z\leq 1.4$ and for galaxies at $1.4 < z \leq 2.7$ ($2.7 < z \leq 3.4$) with masses above
 $10^{10.45}~M_{\odot}$ ($10^{10.66}~M_{\odot}$).
For galaxies below these respective cutoff masses at $1.4 < z \leq 2.7$ and $2.7 < z \leq 3.4$,
 and for all galaxies at $z>3.4$, we assumed $Z_*=0.0031$ (0.27~$Z_{\odot}$) and the SMC
 extinction curve of \citet{gor2003}.
Before fitting, the contribution of nebular emission lines to the photometric flux densities was removed
 based on line fluxes measured from the NIRSpec spectra, using the method described in \citet{san2021}.
The output of the SED fitting includes inferred values and uncertainties for stellar mass,
star-formation rate (SFR(SED)), continuum reddening, and stellar population age.

\subsection{Reddening Correction, SFR, and Line Ratios}\label{sec:sfr}

The amount of dust reddening (\ebvgas) was inferred from the Balmer decrement H$\alpha$/H$\beta$ \citep{sha2023}
 using the Milky Way extinction curve \citep{car1989}, which is consistent with direct constraints
  on the nebular attenuation curve at $z\sim2$ \citep{red2020}.
We assumed an intrinsic ratio of H$\alpha$/H$\beta=2.79$, calculated with \texttt{pyneb} \citep{lur2015}
 for an electron temperature of $T_e=15,000$~K that is typical of high-redshift moderately
 low-metallicity sources \citep[e.g.,][]{san2020,cur2023}.
For targets at $z>6.5$ for which H$\alpha$ was not covered, \ebvgas\ was derived from
 H$\gamma$/H$\beta$ assuming an intrinsic ratio of 0.47.
SFR(H$\alpha$) was derived from the dust-corrected H$\alpha$ luminosity using a metallicity-dependent
 conversion factor based on BPASS population synthesis models including binary effects,
 where galaxies in the 1.4~$Z_{\odot}$+Calzetti case above use a $Z_*=0.02$
 conversion factor of $10^{-41.37} (M_{\odot}{\rm yr}^{-1})/(\mbox{erg~s}^{-1})$, while
 those in the 0.27~$Z_{\odot}$+SMC case use a $Z_*=0.001$ conversion factor of
 $10^{-41.67} (M_{\odot}{\rm yr}^{-1})/(\mbox{erg~s}^{-1})$.
For $z>6.5$ galaxies, we instead infer SFR(H$\beta$) from the dust-corrected H$\beta$ luminosity
using the low-metallicity conversion factor assuming H$\alpha$/H$\beta=2.79$.

Emission-line ratios were calculated from the measured line fluxes.
Ratios of closely-spaced lines (e.g., [N\ii]/H$\alpha$, [O\iii]/H$\beta$, [S\ii]/H$\alpha$, [O\one]/$\alpha$)
 were not corrected for dust reddening and measurement of these ratios thus has no requirement
 of a robust \ebvgas\ constraint.
For ratios involving lines significantly separated in wavelength (e.g., \ott, \rtt), line fluxes
 were first corrected for dust attenuation using the \ebvgas\ values derived above and assuming the \citet{car1989} curve.
As such, \ott\ and \rtt\ were only measured for sources with S/N$\ge$3 detections of
 either H$\alpha$ and H$\beta$, or H$\beta$ and H$\gamma$, for robust \ebvgas\ constraints.
Uncertainties on reddening, SFR, and line ratios were estimated from the interior 68th-percentile
 of distributions derived by perturbing the observed line fluxes by
 their uncertainties, rederiving \ebvgas, and recalculating SFR and all line ratios 5000 times.

\subsection{Sample selection}

For this analysis, we began by selecting the 172 galaxies with robust spectroscopic redshifts at $z\ge2.0$.
Eight targets were removed from the sample as candidate AGN based on the presence of broad emission around bright lines or measured [N\ii]/H$\alpha>0.5$.
We divided the resulting sample of 164 galaxies into 5 bins in redshift containing roughly equal number of sources, and refer to these bins based on the median redshift of sources in each.
The sample analyzed in this work has 29 galaxies at $2.0\le z<2.7$ ($z\sim2.3$); 29 galaxies at $2.7\le z<4.0$ ($z\sim3.3$); 36 galaxies at $4.0\le z<5.0$ ($z\sim4.5$); 43 galaxies at $5.0\le z<6.5$ ($z\sim5.6$); and 27 galaxies at $6.5\le z<9.3$ ($z\sim7.5$).
The full suite of rest-optical lines ([O\ii] to [S\ii]) is covered for the lower four redshift bins, while the $z>6.5$ bin lacks coverage of the redder rest-optical lines ([O\one], H$\alpha$, [N\ii], and [S\ii]).
In each line ratio vs.\ line ratio diagram, we plot galaxies with S/N$\ge$3 for the relevant lines for both ratios as detections, while sources with a non-detection that provides a one-sided limit in one ratio while the lines of the other ratio are detected are ploted as $3\sigma$ limits.

\subsection{Composite Spectra}\label{sec:stacks}

\begin{figure*}
\centering
\includegraphics[width=\textwidth]{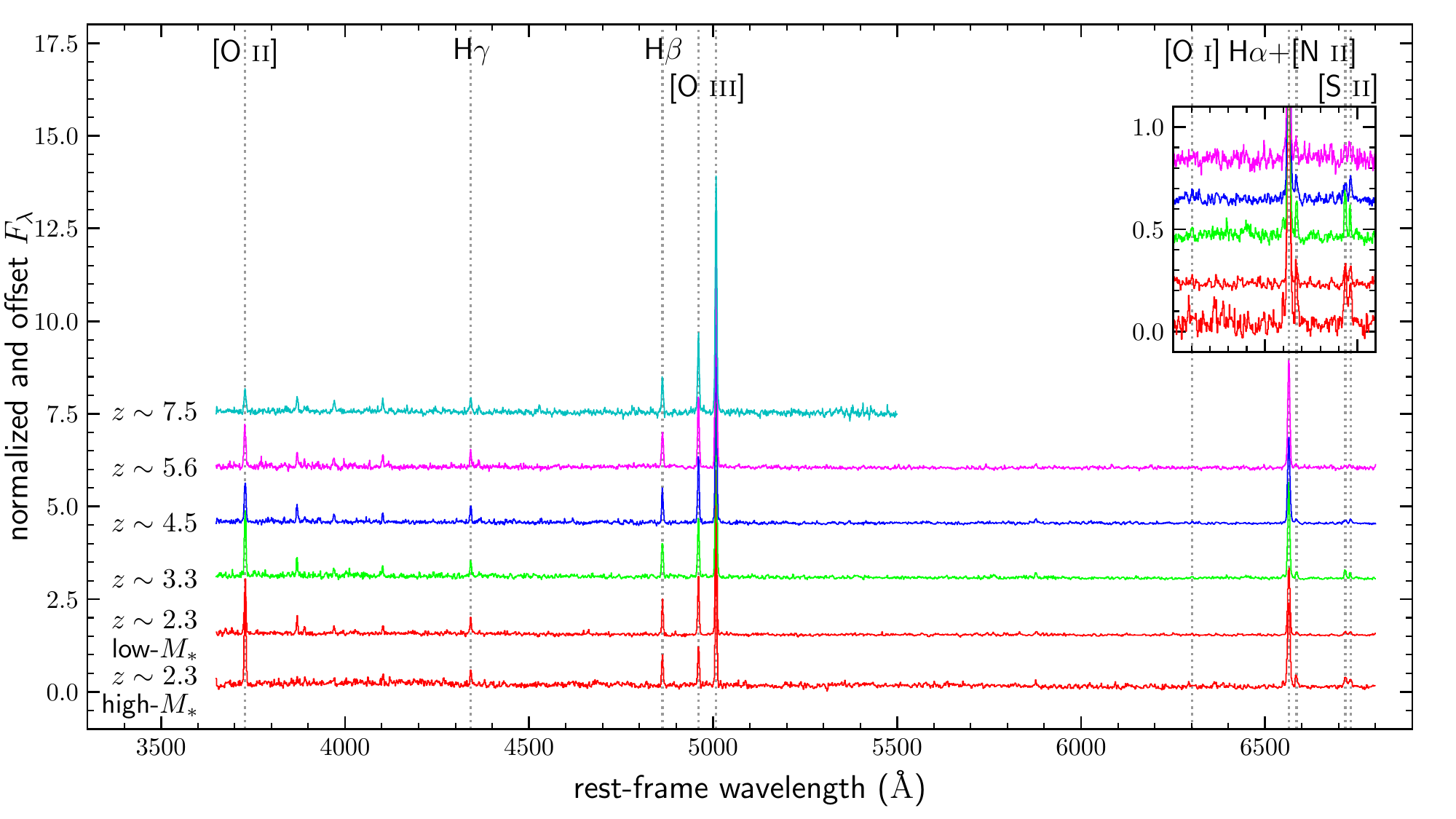}
\caption{Composite spectra of samples in each redshift interval, constructed as described in Sec.~\ref{sec:stacks}.
Each spectrum has been normalized to the measured H$\beta$ intensity and offset vertically for display purposes.
The inset panel zooms in on the the weak red rest-optical lines for the $z<6.5$ composites on the same scale, vertically offset for clarity.
The measured line ratios and uncertainties are reported in Table~\ref{tab:stacks}.
}\label{fig:stacks}
\end{figure*}

Weak lines such as [N\ii], [S\ii], and [O\one] are not detected in many of the individual spectra.
We construct composite spectra in different redshift bins using the following method to assess sample-averaged line ratios including information from galaxies for which all necessary lines are not detected.
For each object in a bin, the 1D spectra from each grating configuration are first normalized by the measured H$\alpha$ flux and shifted to the rest frame.
The spectra are then linearly interpolated onto a uniform wavelength grid with spacing equal to the minimum rest-frame wavelength sampling of all targets in the bin determined by the G140M sampling of the highest-redshift source in the bin.
The 1D spectra in each grating configuration are then combined into a single 1D spectrum, where the inverse-variance weighted mean is used to combine values at wavelengths where there is overlap between two neighboring gratings.
Once these steps have been completed for all targets, the composite spectrum is computed by taking the median value of the individual combined 1D spectra at each wavelength.
We also compute a composite SED model by median-combining the individual best-fit SED models after normalizing by H$\alpha$ flux and shifting to the rest frame.

Line intensities are measured from the normalized composite spectrum using the same methods described in Sec.~\ref{sec:linefitting}, using the composite SED as the continuum model to correct for Balmer absorption.
The dust correction is achieved using the H$\alpha$/H$\beta$ ratio measured from the composite, and uncorrected and dust-corrected line ratios are calculated.
Dust-corrected SFR(H$\alpha$) is calculated by muliplying the dust-corrected, normalized H$\alpha$ intensity by the median H$\alpha$ flux of the individual objects, converting to luminosity using the median redshift of the individual targets, and assuming the low-metallicity conversion factor described in Sec.~\ref{sec:sfr}.
Uncertainties on all properties derived from composite spectra are estimated by bootstrap resampling the targets with replacement, perturbing the individual spectra according to the error spectra, repeating the stacking process and remeasuring all line intensities and derived quantities.
This process was repeated 500 times, and the uncertainty on each quantity was taken to be the interior 68th-percentile bounds on the resulting distributions.
The composite uncertainties thus represent the combination of both measurement uncertainty and sample variance.
Because of differences in rest-frame wavelength coverage due to redshift and position on the slitmask, chip gaps, and masked regions in the 1D spectra, every galaxy in a bin does not contribute to the composite at every wavelength.
However, over the wavelength range used in this analysis, $>85\%$ of the galaxies included in each composite contribute at each wavelength such that the resulting composite represents an unbiased sample-average of the individual sources.
The detector chip gap covers $\approx10\%$ of the total accessible wavelength range in each medium-resolution grating configuration, such that the deficit in total galaxies at any particular wavelength in this range is predominantly due to the lack of spectral information in the chip gap.

\begin{table*}
 \centering
 \caption{Properties of composite spectra for the CEERS/NIRSpec targets at $z=2.0-9.3$.
 }\label{tab:stacks}
 \setlength{\tabcolsep}{2.5pt}
 \renewcommand{\arraystretch}{1.5}
 \begin{tabular}{ l l r r r r r r r r r }
   \hline\hline
   sample  &
   $N_{\text{gal}}$$^a$ &
   $z_{\text{med}}$  &
   $\log{\left(\frac{M_*}{\mbox{M}_\odot}\right)}$$^b$ &
 log$\left(\frac{\text{SFR(H}\alpha)}{\text{M}_\odot/\text{yr}}\right)$ &
 log$\left(\frac{[\text{O}\ \textsc{iii}]}{\text{H}\beta}\right)$ &
 log$\left(\frac{[\text{N}\ \textsc{ii}]}{\text{H}\alpha}\right)$ &
 log$\left(\frac{[\text{S}\ \textsc{ii}]}{\text{H}\alpha}\right)$ &
 log$\left(\frac{[\text{O}\ \textsc{i}]}{\text{H}\alpha}\right)$ &
 log$\left(\text{R}_{32}\right)$ &
 log$\left(\text{O}_{32}\right)$  \\
   \hline\hline
\parbox[c]{1.5cm}{$z\sim2.3$\\(high-$M_*$)}  &  13  &  2.337  &  $9.78^{+0.05}_{-0.01}$  &  $0.73^{+0.03}_{-0.08}$  &  $0.58^{+0.02}_{-0.03}$  &  $-0.93^{+0.04}_{-0.11}$  &  $-0.68^{+0.04}_{-0.03}$  &  $<-1.52$  &  $0.94^{+0.03}_{-0.04}$  &  $-0.03^{+0.04}_{-0.02}$    \\
\parbox[c]{1.5cm}{$z\sim2.3$\\(low-$M_*$)}  &  13  &  2.302  &  $9.09^{+0.04}_{-0.07}$  &  $0.32^{+0.04}_{-0.04}$  &  $0.73^{+0.02}_{-0.02}$  &  $-1.39^{+0.08}_{-0.06}$  &  $-0.97^{+0.04}_{-0.05}$  &  $<-1.87$  &  $0.95^{+0.02}_{-0.02}$  &  $0.45^{+0.03}_{-0.02}$    \\
$z\sim3.3$  &  27  &  3.321  &  $9.64^{+0.04}_{-0.08}$  &  $0.61^{+0.06}_{-0.04}$  &  $0.68^{+0.02}_{-0.01}$  &  $-1.18^{+0.03}_{-0.06}$  &  $-0.92^{+0.04}_{-0.05}$  &  $-1.75^{+0.15}_{-0.12}$  &  $0.91^{+0.03}_{-0.01}$  &  $0.40^{+0.02}_{-0.04}$    \\
$z\sim4.5$  &  31  &  4.560  &  $9.41^{+0.12}_{-0.06}$  &  $0.75^{+0.07}_{-0.04}$  &  $0.78^{+0.03}_{-0.01}$  &  $-1.29^{+0.05}_{-0.07}$  &  $-1.08^{+0.05}_{-0.05}$  &  $<-1.80$  &  $0.98^{+0.04}_{-0.01}$  &  $0.49^{+0.03}_{-0.04}$    \\
$z\sim5.6$  &  38  &  5.507  &  $8.57^{+0.04}_{-0.13}$  &  $0.67^{+0.04}_{-0.05}$  &  $0.75^{+0.02}_{-0.02}$  &  $-1.46^{+0.12}_{-0.09}$  &  $-1.30^{+0.17}_{-0.02}$  &  $<-1.67$  &  $0.95^{+0.01}_{-0.02}$  &  $0.52^{+0.07}_{-0.01}$    \\
$z\sim7.5$  &  24  &  7.461  &  ---  &  ---  &  $0.81^{+0.03}_{-0.02}$  &  ---  &  ---  &  ---  &  $0.97^{+0.03}_{-0.02}$  &  $0.94^{+0.10}_{-0.09}$    \\
   \hline\hline
 \end{tabular}
 \begin{flushleft}
 $^{a}$ {Number of galaxies in each composite.}
 $^{b}$ {Median stellar mass of galaxies in each composite.}
 \end{flushleft}
\end{table*}

The $z\sim2.3$ spectra are of sufficient depth to detect [N\ii] and [S\ii] after dividing the sample into two bins.
We choose to divide the $z\sim2.3$ sample into two bins at the median stellar mass, and consequently
 require S/N$\ge$3 for H$\alpha$ and a measured stellar mass from SED fitting, resulting in a $z\sim2.3$ stacking sample of 26 out of the 29 targets at this redshift.
For the higher redshift intervals, [N\ii] and [S\ii] were not detected in all composites when dividing into two stellar mass bins.
Accordingly, we stack all $z\ge2.7$ galaxies in a single bin per redshift interval to achieve sufficient S/N for the weak lines, requiring only that H$\alpha$ has S/N$\ge$3 for normalization.
This selection results in stacking samples of 27 out of 29 galaxies at $z\sim3.3$, 31 out of 36 galaxies at $z\sim4.5$, and 38 out of 43 galaxies at $z\sim5.6$.
Since H$\alpha$ is not covered at $z>6.5$, we instead use [O\iii]$\lambda$5008 normalization throughout the stacking process, the composite H$\gamma$/H$\beta$ ratio for dust correction, and SFR(H$\beta$) in this highest redshift bin.
The requirement of [O\iii]$\lambda$5008 S/N$\ge$3 yields a stacking sample of 24 out of 27 galaxies at $z\sim7.5$.

The composite spectra in each of these bins are displayed in Figure~\ref{fig:stacks}, showing significant detections of [O\ii], H$\beta$, and [O\iii]$\lambda\lambda$4960,5008 in all bins, as well as H$\alpha$, [N\ii]$\lambda$6585, and [S\ii]$\lambda\lambda$6718,6733 in the $z<6.5$ bins.
Ratios of these lines are the subject of this analysis, and are reported in Table~\ref{tab:stacks} alongside stellar mass, SFR, and redshift.
Many weaker lines including [Ne\iii]$\lambda$3870, higher-order Balmer lines, [O\iii]$\lambda$4363, and He~\one$\lambda$5877 are also present and will be the subject of future studies.

\subsection{Low-redshift Comparison Samples}

We compared to $z\sim0$ star-forming galaxies selected from the Sloan Digital Sky Survey \citep[SDSS][]{yor2000} using the selection criterion of \citet{and2013}, resulting in a sample of $\approx$200,000 star-forming galaxies at $z\sim0.07$.
We use emission-line measurements and stellar masses from the MPA-JHU catalogs\footnote{\url{https://wwwmpa.mpa-garching.mpg.de/SDSS/DR7/}} and calculate reddening-corrected line ratios and SFR(H$\alpha$) as described above, adopting the solar-metallicity SFR(H$\alpha$) conversion factor (Sec.~\ref{sec:sfr}).
We also compare to the collection of $\sim1000$ $z=0$ H\ii\ regions presented in \citet{san2017}, an expansion of the catalog of \citet{pil2016}.
Line ratios are calculated from the dust-corrected line intensities tabulated in this catalog.

\section{Results}\label{sec:results}

\subsection{Sample Properties}\label{sec:sfrmstar}

\begin{figure}
\centering
\includegraphics[width=0.95\columnwidth]{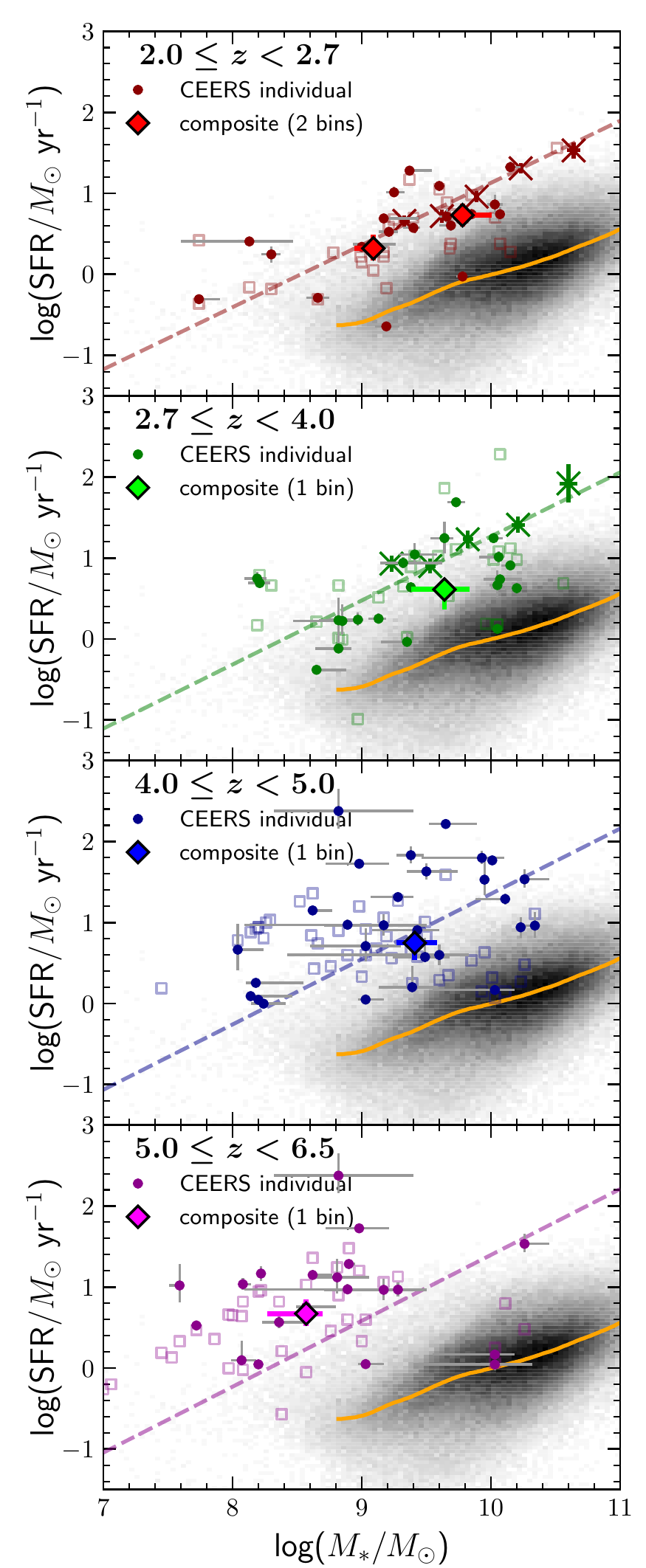}
\caption{SFR vs.\ \mstar\ for samples at $z=2.0-6.5$.
Filled circles with error bars display SFR(H$\alpha$), while hollow squares present SFR(SED).
In each panel, the dashed line displays the star-forming main sequence from the parameterization of \citet{spe2014}, calculated at the median redshift of the galaxies in each bin.
The gray two-dimensional histogram displays the distribution of $z\sim0$ star-forming galaxies from SDSS, for which the orange line shows the median trend.
}\label{fig:sfrmstar}
\end{figure}

We begin by investigating the sample properties of the galaxies in each redshift interval.
Before analyzing the excitation properties of the sample, it is of interest to understand whether these galaxies lie along the ``main sequence'' of star formation in the SFR vs.\ \mstar\ plane because gas-phase metallicity and ionization state have been found to depend on SFR and/or specific SFR (sSFR=SFR/\mstar) \citep[e.g.,][]{man2010,san2016,san2021,kas2019}.
Figure~\ref{fig:sfrmstar} shows SFR vs.\ \mstar\ for the individual galaxies and composites at $z=2.0-6.5$.
For individual galaxies, both SFR(H$\alpha$) (filled circles) and SFR(SED) (hollow squares) are shown.
SFR and \mstar\ display positive correlations at each redshift, with H$\alpha$- and SED-derived SFRs showing good agreement on average.
We compare these measurements to the parameterized star-forming main sequence of \citet{spe2014} evaluated at the median redshift of each bin.
For the lowest two redshift intervals, we also compare to Balmer-line SFR measurements from stacked spectra of star-forming galaxies from the MOSDEF survey at $z\sim2.3$ and $z\sim3.3$ from \citet{san2021}.
Both the \citet{spe2014} relation and the MOSDEF stacks have been adjusted 0.34~dex lower in SFR to account for the difference between the solar-metallicity \citet{hao2011} SFR conversion factor on which those works are based and the low-metallicity BPASS conversion factor that is predominantly used here.

The CEERS/NIRSpec samples at $z\sim2.3$, $z\sim3.3$, and $z\sim4.5$ are representative of main sequence galaxies at these redshifts, with individual targets scattering about the mean relation while the composites show good agreement with the \citet{spe2014} relation.
The $z\sim2.3$ and $z\sim3.3$ CEERS targets further display excellent agreement with the main-sequence MOSDEF samples matched in redshift.
At $z\sim5.6$, the CEERS sample lies mostly above the main sequence, with the composite having 0.4~dex higher SFR at fixed \mstar.
This offset is likely the result of the lower average \mstar\ of the $z\sim5.6$ sample combined with the higher limiting line luminosity at this higher redshift, relative to the $z<5.0$ samples.
We thus find that the CEERS/NIRSpec targets at $z=2.0-5.0$ are representative of the main-sequence star-forming galaxy population, while the $z\sim5.6$ sample is moderately biased toward high sSFR.
The $z\sim7.5$ sample was excluded from this analysis due to the lack of a robust determination of the main sequence at such high redshifts (the \citet{spe2014} analysis only included data up to $z\sim6$).
However, it is likely that the $z\sim7.5$ is also biased toward starbursts like the $z\sim5.6$ sample.

\subsection{The BPT Diagrams at $z=2.0-6.5$}\label{sec:bpt}

\begin{figure*}
\centering
\includegraphics[width=0.95\textwidth]{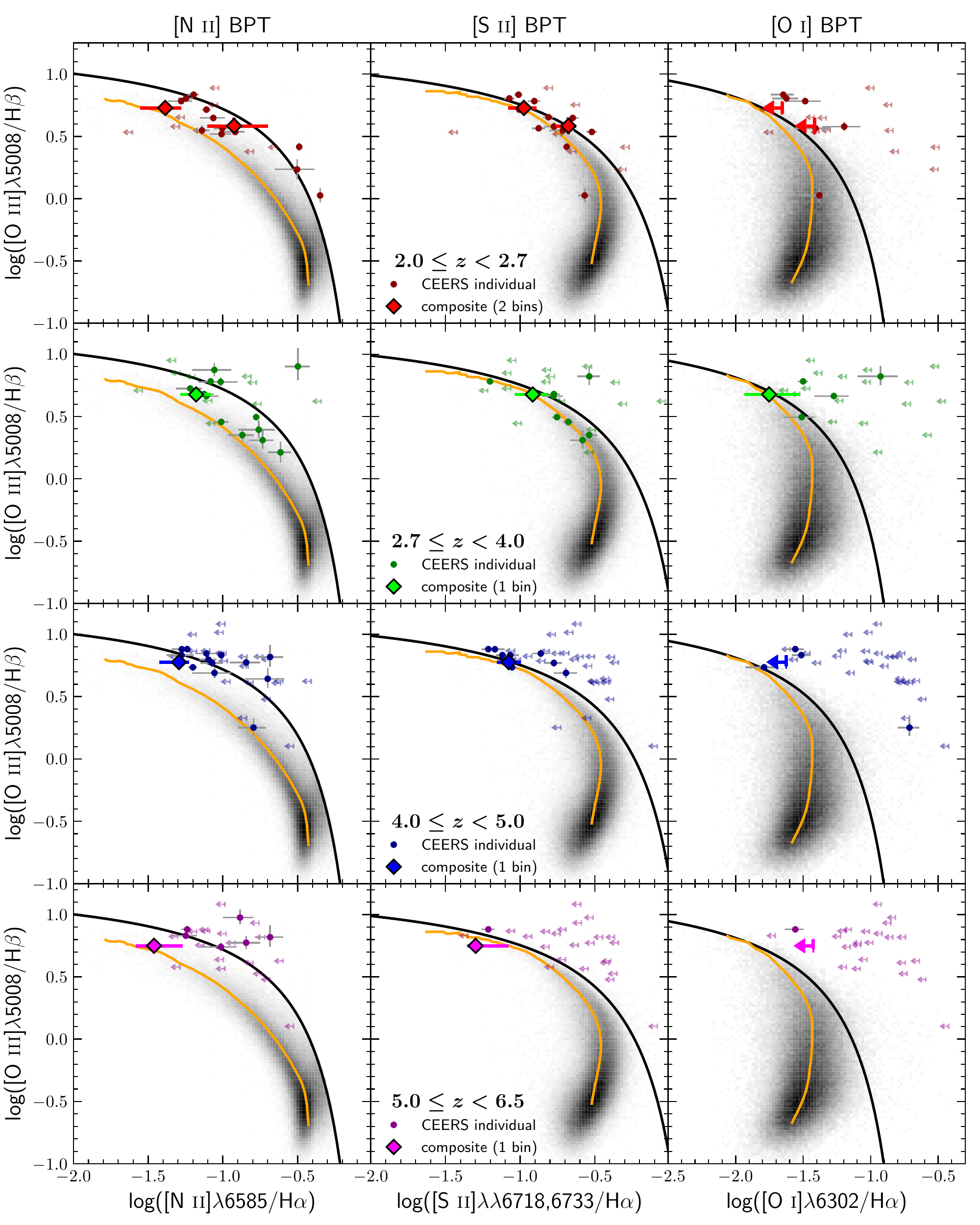}
\caption{The [N\ii] (left column), [S\ii] (middle column), and [O\one] (right column) BPT diagrams at $z=2.0-6.5$.
Each row displays individual galaxies and composite spectra for a single redshift interval.
Arrows denote 3$\sigma$ upper limits when the weakest line was not detected.
The gray two-dimensional histogram shows the distribution of $z\sim0$ star-forming galaxies from SDSS, while the orange line provides the median trend of this sample.
The black solid line displays the local empirical demarcation between star-forming galaxies and AGN from \citet{kau2003} ([N\ii] BPT) or \citet{kew2006} ([S\ii] and [O\one] BPT diagrams).
}\label{fig:bptz}
\end{figure*}

We now turn to the excitation properties of the CEERS/NIRSpec galaxies as revealed by their emission-line ratios.
Figure~\ref{fig:bptz} displays the $z=2.0-6.5$ samples and composite spectra in the [N\ii], [S\ii], and [O\one] BPT diagrams separately for each redshift interval, alongside the $z\sim0$ SDSS star-forming galaxies.
At $z\sim2.3$, the CEERS targets display decreasing \othb\ with increasing \ntha\ along a sequence that is offset from the $z\sim0$ locus toward higher \ntha\ at fixed \othb\ (or, equivalently, higher \othb\ at fixed \ntha).
\othb\ also decreases with increasing \stha, with the $z\sim2.3$ galaxies lying along the $z\sim0$ locus.
These trends are in agreement with past studies at $z\sim2.3$ using ground-based spectroscopy of hundreds of galaxies \citep[e.g.,][]{sha2015,sha2019,ste2014,str2017,str2018,san2016}.
While [O\one]$\lambda$6302 is not significantly detected in either $z\sim2.3$ composite, 3$\sigma$ upper limits suggest that $z\sim2.3$ galaxies lie below the \citet{kew2006} demarcation between local star-forming galaxies and AGN in the [O\one] BPT diagram.

Before the advent of {\it JWST}, it was not possible to measure the lines required for BPT diagrams at $z>2.7$.
We now present the first statistical results on the position of star-forming galaxies in these diagnostic excitation diagrams at $z=2.7-6.5$.
Galaxies at $z\sim3.3$ also display a clear trend of decreasing \othb\ with increasing \ntha\ and \stha.
At $z\sim4.5$ and $z\sim5.6$, such trends cannot be reliably assessed with the current sample due to the large number of upper limits on [N\ii] and/or [S\ii] and the inability to detect these lines in stacks of more than a single bin, motivating the collection of a considerably larger statistical sample to understand the shape of the BPT excitation sequences at these redshifts.
However, the single-bin stacks allow us to assess average offsets relative to the local BPT sequences.
Galaxies at $z\sim3.3$, $z\sim4.5$, and $z\sim5.6$ are all offset above the [N\ii] BPT sequence on average, similar to what has been found at $z\sim2.3$, while the composites are consistent with no offset relative to $z\sim0$ in the [S\ii] BPT diagram at each of these redshifts.
The $z\sim3.3$ composite further displays no offset in the [O\one] BPT diagram.
In all $z>2$ samples, individual galaxies with no evidence of significant AGN emission scatter above the local demarcations between AGN and star-forming galaxies derived by \citet{kau2003} and \citet{kew2006}.
While reliable alternative AGN identification information (e.g., X-ray, rest-near-IR colors) are not as readily available at $z>4$, this result suggests that the $z\sim0$ BPT diagram demarcations are not reliable separators of star-forming galaxies and AGN in the high-redshift universe, consistent with earlier findings at $z\sim2.3$ \citep{coi2015}.

\begin{figure*}
\centering
\includegraphics[width=\textwidth]{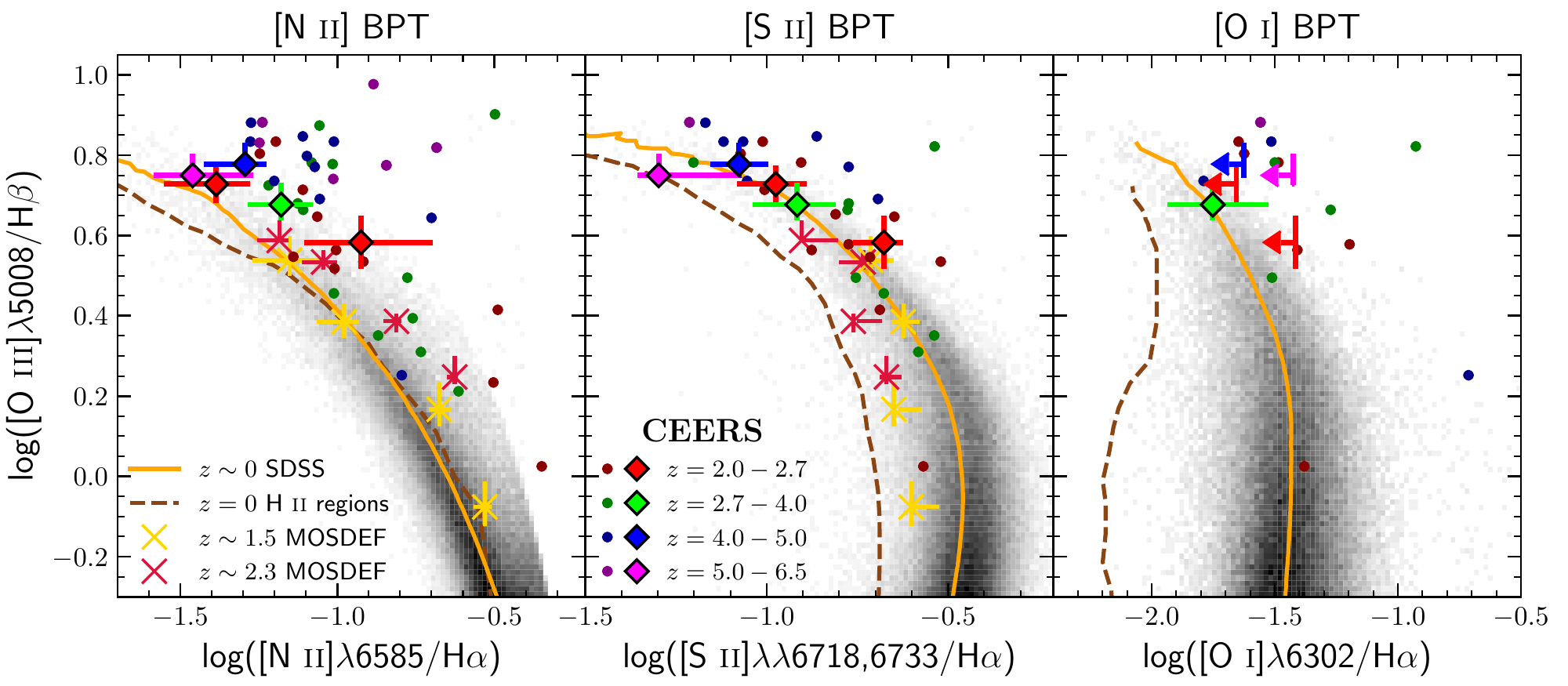}
\caption{The [N\ii] (left), [S\ii] (middle), and [O\one] (right) BPT diagrams for samples at $z=0-6.5$.
Composite spectra and individually-detected CEERS galaxies are displayed as in Fig.~\ref{fig:bptz}.
Composite spectra of star-forming galaxies at $z\sim1.5$ and $z\sim2.3$ from the MOSDEF survey are displayed as "X"s \citep{sha2019}.
The distribution of $z\sim0$ star-forming galaxies from SDSS is shown in the gray two-dimensional histogram, with the median trend provided in orange.
The brown dashed line denotes the median line ratio sequences derived from the catalog of $\sim1000$ $z=0$ H\ii\ regions presented in \citet{san2017} \citep[see also][]{pil2016}.
}\label{fig:bptall}
\end{figure*}

Figure~\ref{fig:bptall} compares individual detections and composites of the CEERS $z=2.0-6.5$ samples to $z\sim0$ star-forming galaxies and H\ii\ regions, as well as ground-based composites from the MOSDEF survey in the [N\ii] and [S\ii] BPT diagrams \citep{sha2015}.
As shown by \citet{sha2015}, the MOSDEF samples reveal a steady evolution of the BPT excitation sequence toward higher \othb\ and \ntha\ from $z\sim0$ to $z\sim1.5$ to $z\sim2.3$.
The $z\sim2.3$ CEERS stacks are consistent with the $z\sim2.3$ MOSDEF sequence, though the CEERS data extend to lower \ntha\ as a consequence of its lower \mstar\ range, which  includes lower-metallicity galaxies than those in MOSDEF.
The $z>4.0$ samples in particular lie predominantly at lower \mstar\ and \ntha\ than MOSDEF.
We consider the combined $z\sim2.3$ CEERS+MOSDEF composites as defining a single excitation sequence at this redshift in which the low-\mstar\ CEERS composite extends the MOSDEF sequence to lower \ntha.
The $z\sim3.3$, $z\sim4.5$, and $z\sim5.6$ composites are all consistent with this combined $z\sim2.3$ sequence within the uncertainties, suggesting no strong evolution of the [N\ii] BPT excitation sequence at redshifts higher than $z\sim2.3$ within the precision limits of the current samples.

In the [S\ii] BPT diagram, we again find that the CEERS samples at $z=2.0-6.5$ appear to lie on the same excitation sequence.
This high-redshift [S\ii] sequence is consistent with the locus of low-metallicity $z\sim0$ galaxies from SDSS but significantly offset toward higher \stha\ at fixed \othb\ relative to $z=0$ H\ii\ regions, similar to findings at $z\sim2$ by \citet{sha2019}.
In the [O\one] BPT diagram, the composite detection at $z\sim3.3$ yields a similar result, being consistent with the $z\sim0$ star-forming galaxies but offset from the local H\ii\ regions.

\subsection{The \ott$-$\rtt\ Diagram}\label{sec:r23o32}

\begin{figure}
\centering
\includegraphics[width=0.85\columnwidth]{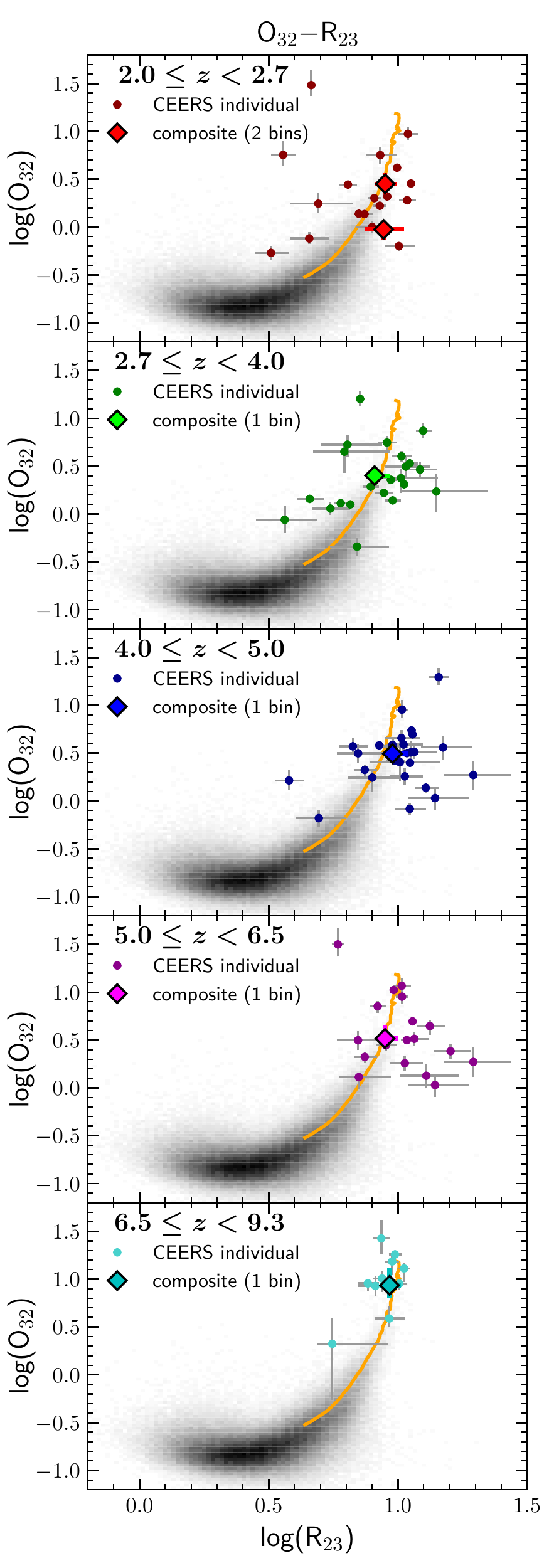}
\caption{The \ott\ vs.\ \rtt\ diagram at $z=2.0-9.3$.
Both line ratios have been corrected for dust reddening.
Each row displays individual galaxies and composite spectra for a single redshift interval.
The gray two-dimensional histogram and orange line shows $z\sim0$ star-forming galaxies as in previous figures.
}\label{fig:r23o32z}
\end{figure}

\begin{figure}
\centering
\includegraphics[width=\columnwidth]{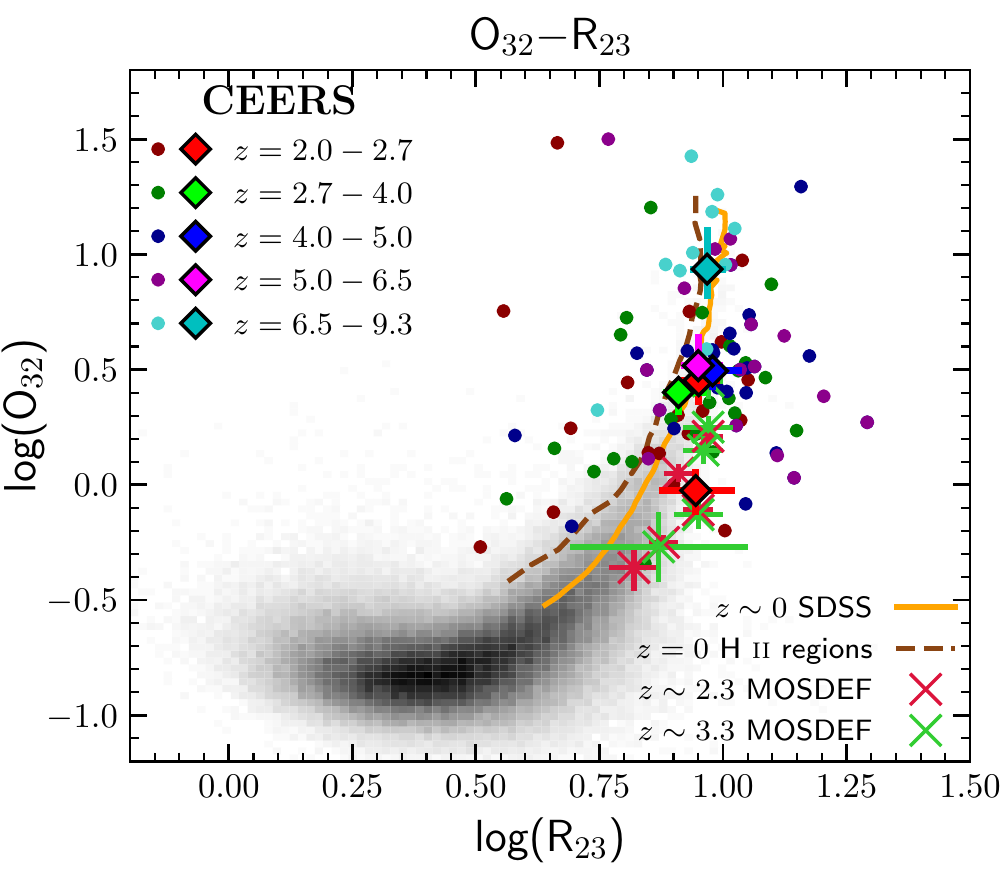}
\caption{
The \ott\ vs.\ \rtt\ diagram for samples at $z=0-9.3$.
Composite spectra and individually-detected CEERS galaxies are displayed as in Fig.~\ref{fig:r23o32z}.
Composite spectra of star-forming galaxies at $z\sim2.3$ and $z\sim3.3$ from the MOSDEF survey are displayed as "X"s \citep{san2021}.
The distribution of $z\sim0$ star-forming galaxies from SDSS is shown in the gray two-dimensional histogram, with the median trend provided in orange.
The brown dashed line denotes the median sequence $z=0$ H\ii\ regions.
}\label{fig:r23o32all}
\end{figure}

The position of CEERS galaxies and composites at $z=2.0-9.3$ in the \ott\ vs.\ \rtt\ diagram is shown separately for each redshift interval in Figure~\ref{fig:r23o32z}, now including the $z\sim7.5$ sample for which the necessary lines in this diagram are still accessible with NIRSpec.
At each redshift, the CEERS galaxies lie near the excitation sequence described by the median relation of the $z\sim0$ SDSS star-forming galaxies, though a non-negligible fraction of high-redshift sources have very high \rtt$>$10 where local star-forming sources are exceedingly rare.
Figure~\ref{fig:r23o32all} displays all five high-redshift samples alongside local samples and stacked measurements from the MOSDEF survey at $z\sim2.3$ and $z\sim3.3$ \citep{san2021}.
We find that $z=2-4$ targets are offset from the $z\sim0$ SDSS sequence toward higher \rtt\ at fixed \ott\ at log(\ott$)\lesssim0.2$, but this offset appears to decrease toward higher \ott\ until the high-redshift samples are coincident with the median $z\sim0$ SDSS sequence.
However, both the SDSS and high-redshift galaxies are offset from the $z=0$ H\ii\ regions toward higher \rtt\ at fixed \ott.
As in the BPT diagrams, no significant evolution is apparent above $z=2.0$, with the composites at $z\sim2.3$ (low-\mstar), $z\sim3.3$, $z\sim4.5$, and $z\sim5.6$ in particular falling in a nearly identical position.
Interestingly, the $z\sim7.5$ sample lies at significantly higher \ott\ than the $z<6.5$ samples, with an average \ott$\approx$10.

\section{Discussion}\label{sec:discussion}

The rest-frame optical emission-line ratio measurements from {\it JWST}/NIRSpec presented above enable the first glimpse of the metallicity, excitation, and ionization properties of the typical star-forming galaxy population at $z>4$ based on rest-optical emission lines, and an expanded view at $z=2.7-4$ where rest-optical lines redward of [O\iii]$\lambda$5008 were inaccessible prior to {\it JWST}.
The $z>2.7$ samples uniformly display weak [N\ii] (log(\ntha$)<-1.0$) and [S\ii] (log(\stha$)<-0.8$), and high \othb\ ($\gtrsim3$), \rtt\ ($\gtrsim6$), and \ott\ ($\gtrsim2$).
This pattern of emission-line strengths indicates that low gas-phase metallicities and high degrees of ionization are common among early-universe galaxies, qualitatively consistent with expectations based on the evolution of the mass-metallicity relation \citep{san2021} and the anti-correlation between metallicity and ionization parameter \citep{per2014}.
However, the fact that \othb\ and \rtt\ remain high in all CEERS/NIRSpec samples (including at $6.5\le z<9.3$) suggests that these galaxies are not extremely metal-poor in oxygen (12+log(O/H$)<7.7$ or $<0.1~Z_{\odot}$).
Both \othb\ and \rtt\ are double valued, where these line ratios increase with increasing metallicity at 12+log(O/H$)\lesssim8.0$ (0.2~$Z_{\odot}$), while decreasing at higher metallicities \citep[e.g.,][]{kew2019}.
The uniformly high values of \othb\ and \rtt\ across the CEERS/NIRSpec samples thus imply metallicities of $\sim0.1-0.3$~$Z_{\odot}$ where these line ratios peak in local samples \citep[e.g.,][]{pil2016}.
In contrast, extremely metal-poor local galaxies at $<0.05$~$Z_{\odot}$ display significantly lower \rtt\ and \othb\ \citep[e.g.,][]{izo2012,izo2018}, while the galaxy SMACS-0723-04590 at $z=8.496$ with a direct-method metallicity of 12+log(O/H$)\sim7.0$ (0.02~$Z_{\odot}$) has log(\rtt$)\sim0.6$ \citep{cur2023}, lower than nearly all CEERS/NIRSpec galaxies at $z=2-9.3$.
However, the relations between these strong-line ratios and gas-phase metallicity is not yet well-constrained at $z>2$ \citep{san2020}.
A significantly expanded sample of high-redshift galaxies with robust direct-method metallicity measurements is required to accurately translate the set of line ratios presented here into quantitative metallicity and ionization parameter constraints.

It is notable that, in all of the high-redshift intervals, the detection rate of all lines required to derive \rtt\ and \ott\ is much higher than for [N\ii] and [S\ii].
As such, at $z>2$, the bluer rest-optical lines ([O\ii], H$\beta$, and [O\iii]) appear to present a significantly more viable route to metallicity and ionization information than indicators based on [N\ii] or [S\ii], while also having the advantage of being accessible at $z>6.5$ and far into the reionization epoch.
As noted above, the CEERS/NIRSpec sample at $z=6.5-9.3$ displays significantly higher \ott\ on average than the sample at $z\sim5.6$, while the $z\sim5.6$ sample has similar \ott\ to the $z=2.7-5$ galaxies.
Extreme levels of ionization are thus common among the CEERS/NIRSpec $z>6.5$ sample, but it is not clear whether this result indicates such high ionization is common among all reionization-era galaxies since we do not know whether the CEERS $z>6.5$ sources are representative of the star-forming population.
Extreme \ott\ values may indicate the presence of density-bounded H\ii\ regions that enable significant escape of Lyman-continuum (LyC) photons \citep{nak2014}.
However, results at lower redshift suggest high \ott\ may be a necessary but not sufficient condition for significant LyC escape and similarly high \ott\ values can be produced in models with no LyC escape \citep[e.g.,][]{sta2015,tan2019,izo2021,flu2022}.
The seemingly rapid increase in the degree of ionization between $z\sim5.6$ and $z>6.5$ among the CEERS galaxies thus does not necessarily indicate an increase in LyC escape.

A key result is that all CEERS/NIRSpec samples spanning $z=2.0-6.5$ fall along the same sequence in the [N\ii] and [S\ii] BPT diagrams, as well as in the \ott-\rtt\ diagram.
Specifically, the $z\sim3.3$, $z\sim4.5$, and $z\sim5.6$ composites all have line ratios that are consistent within the uncertainties with the low-\mstar\ $z\sim2.3$ composite in all line ratio diagrams presented here.
The similar line ratio properties of these samples suggest that the ionization conditions of gas in H\ii\ regions do not significantly evolve between $z=2$ and $z=6.5$, spanning 2.4~Gyr of cosmic history.
If this result is confirmed with expanded future samples, then one important implication is that the same diagnostic calibrations can be used to translate strong-line ratios into physical properties such as metallicity and ionization parameter across this wide redshift range.

We find that the unified $z>2$ excitation sequence is offset toward higher \ntha\ at fixed \othb\ relative to $z\sim0$ SDSS star-forming galaxies, but lies on the $z\sim0$ locus in the [S\ii] BPT diagram (and the [O\one] BPT diagram for the $z\sim3.3$ composite).
This comparison is complicated by the presence of diffuse ionized gas (DIG) emission in the galaxy-integrated SDSS fiber spectra, significantly affecting galaxy-integrated line ratios \citep{zha2017,san2017,val2019,bel2022}.
\citet{sha2019} argued that the high-SFR nature of high-redshift galaxies suggests negligible DIG emission in their integrated spectra, such that a comparison should instead be made to local H\ii\ regions or SDSS spectra that have been corrected for DIG contamination.
It can be seen in Figures~\ref{fig:bptall} and~\ref{fig:r23o32all} that the CEERS/NIRSpec samples are in fact offset from the median sequence of $z=0$ H\ii\ regions in each of these line ratio diagrams, with higher \ntha, \stha, and \oiha\ at fixed \othb\ and larger \rtt\ at fixed \ott.

These offsets with respect to $z=0$ H\ii\ regions are consistent with high-redshift galaxies uniformly having a harder ionizing spectrum at fixed O/H, driving larger \othb, \ntha, \stha, and \rtt\ \citep[e.g.,][]{ste2016}.
At $z\sim2-3.5$, recent studies have shown that massive stars in these galaxies have chemical compositions with elevated $\alpha$/Fe ratios of $\text{O/Fe}\approx2-5\times\text{O/Fe}_{\odot}$
\citep{ste2016,str2018,str2022,top2020a,top2020b,san2020,cul2021,red2022}.
This $\alpha$-enhancement and Fe-deficiency is a consequence of the short formation timescales of high-redshift galaxies, in which enrichment is dominated by short-timescale core-collapse supernovae while time-delayed Type Ia supernovae have not yet significantly enriched the ISM in Fe.

Our results are thus consistent with $\alpha$-enhanced massive stars driving offsets between local and $z=2.0-6.5$ samples in the BPT and \ott-\rtt\ diagrams, as expected if galaxies are typically younger at higher redshifts.
However, such a trend may be expected to lead to higher levels of $\alpha$-enhancement at higher redshifts, and thus a steadily increasing excitation sequence offset above $z=2$ for which we do not find evidence.
It may be that the theoretical upper-limit on $\alpha$-enhancement from pure core-collapse supernovae enrichment ($\approx5\times\text{O/Fe}_{\odot}$; \citealt{kob2020}) is reached at typical galaxy ages at $\sim2-4$ such that younger stellar populations at $z\sim4-6.5$ do not lead to higher $\alpha$/Fe, or that the precision enabled by the current $z>2.7$ sample sizes ($\sim$30 galaxies per redshift interval) is not great enough to resolve the shift in line ratios based on evolution from $\approx3\times\text{O/Fe}_{\odot}$ at $z\sim2$ to $5\times\text{O/Fe}_{\odot}$ at some higher redshift.
The present sample is also too small to detect weak lines ([N\ii], [S\ii], [O\one]) in multiple bins at $z>2.7$ to test whether the shape of these excitation sequences remains the same over $z=2-6.5$.
An order-of-magnitude larger sample is ultimately needed to approach the level of statistical power reached by ground-based samples at $z\sim2.3$ \citep[e.g.,][]{ste2014,sha2019} and establish the shape and relative offset of $z>2.7$ excitation sequences with high precision.
{\it JWST}/NIRSpec has the capability to provide such a sample and transform our understanding of the chemical and ionization properties of early galaxies in the coming years.

\begin{acknowledgments}
This work is based on observations made with the NASA/
ESA/CSA James Webb Space Telescope. The data were
obtained from the Mikulski Archive for Space Telescopes at
the Space Telescope Science Institute, which is operated by the
Association of Universities for Research in Astronomy, Inc.,
under NASA contract NAS5-03127 for {\it JWST}. 
Support for this work was provided through the NASA Hubble Fellowship
grant \#HST-HF2-51469.001-A awarded by the Space Telescope
Science Institute, which is operated by the Association of Universities
for Research in Astronomy, Incorporated, under NASA contract NAS5-26555.
We also acknowledge support from NASA grant {\it JWST}-GO-01914.
\end{acknowledgments}

\vspace{5mm}
\facilities{{\it JWST}(NIRSpec)}

\bibliography{bpt}{}
\bibliographystyle{aasjournal}

\end{document}